%% file: main.tex
\documentclass[10pt,conference]{IEEEtran}

\input{preamble}

\begin{document}

\date{}

\title{Lessons Learned from a Bare-metal Evaluation of Erasure Coding Algorithms in P2P Networks}
\author{
	\IEEEauthorblockN{Racin Nygaard}
	\IEEEauthorblockA{\textit{Department of Electrical Engineering and Computer Science} \\
		\textit{University of Stavanger}\\
		Stavanger, Norway \\
		racin.nygaard@uis.no}
}

\maketitle

\input{tex/abstract}
\input{tex/background}

\input{tex/ethswarm}
\input{tex/design}
\input{tex/evalutation}
\input{tex/dealingreplication}
\input{tex/dealingsyncing}
\input{tex/conclusion}

\bibliographystyle{plain}
\bibliography{references}
\end{document}

%% file: preamble.tex
\usepackage{tikz}
\usepackage{amsmath}

\usepackage{filecontents}
\usepackage{mdframed}      
\usepackage{xspace}
\usepackage{hyperref}

\newcommand{\figref}[1]{Figure~\ref{#1}\xspace}


%% file: tex/abstract.tex

\begin{abstract}
We have built a bare-metal testbed in order to perform large-scale, reproducible evaluations of erasure coding algorithms. 
Our testbed supports at least 1000 Ethereum Swarm peers running on 30 machines. 
Running experimental evaluation is time-consuming and challenging. 
Researchers must consider the experimental software's limitations and artifacts.
If not controlled, the network behavior may cause inaccurate measurements.
This paper shares the lessons learned from a bare-metal evaluation of erasure coding algorithms and how to create a controlled-environment in a cluster consisting of 1000 Ethereum Swarm peers. 
\end{abstract}

%% file: tex/background.tex

\section{Background and Motivation}
\noindent
We are designing algorithms to protect data in decentralized networks, such as the Ethereum Swarm peer-to-peer network.
The Merkle tree~\cite{Merkle1987digital} is a widely-used hash tree to authenticate data. 
In content-addressed storage, a hash tree can be used to locate the chunks in the network. 
That permits to locate a large number of chunks by knowing only the tree's root. 
Due to the tree structure, it is critical to protect chunks in the path between the root and the data chunks (leaves).

Large-scale experimental evaluation of erasure codes is rarely observed in the literature.
We want to implement algorithms and understand their impact on distributed storage networks. 
Typically, the metrics of interest are latency, throughput, network- and storage overhead.
Many conditions can affect measurements, including individual peers' behavior and the need to run hundreds or thousands of concurrent instances.
To obtain meaningful results, we run our experimental evaluations in a controlled environment and, at the same time, close to a real-world scenario. 
Our bare-metal cluster consists of 30 machines, running up to 1000 Swarm peers~\cite{swarm}.

This paper shares insight from our evaluation environment and lessons learned by observing the peers' behavior.
Our evaluation techniques can help others improve the evaluation of erasure codes in large networks.
We discuss our experimental setup, the main challenges, and the tools developed to solve them.
The source code for the tools will be made available~\cite{snarl-tools}.

%% file: tex/ethswarm.tex

\section{Ethereum Swarm}
\noindent
Swarm is a decentralized storage and communication system.
Swarm's test network is relatively large; a 3-month study found 6,500 unique peers~\cite{kim2018measuring}.
The basic unit of storage is a \emph{chunk} limited to 4K bytes.

Swarm splits a file into chunks for upload and computes a cryptographic hash of each chunk.
This hash is also known as the \emph{content address} and is necessary to access the chunk later.

Chunks that belong to the same file are organized in a Merkle tree.
During retrieval, the client initially queries the network with the root's content address.
It then deciphers the root chunk to retrieve its children's content address, for which it again queries the network.
This process continues until all leaves are retrieved and the original file has been rebuilt.

\begin{figure}
	\centering
	\begin{tikzpicture}[level distance=0.8cm,
	level 1/.style={sibling distance=3cm},
	level 2/.style={sibling distance=0.99cm}]
	\node {Root}
	child {node {I1}
		child {node {L1}}
		child {node {L2}}
		child {node {L3}}
	}
	child {node {I2}
		child {node {L4}}
		child {node {L5}}
		child {node {L6}}
	}
	child {node {I3}
		child {node {L7}}
		child {node {L8}}
		child {node {L9}}
	};
	\end{tikzpicture}
	\caption{Merkle tree illustrated with branching factor 3, resulting in 9 leaves, 3 internal nodes and 1 root.}
\end{figure}
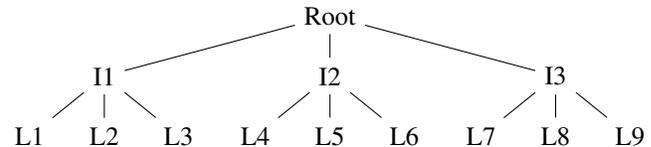

%% file: tex/design.tex

\section{Adding Redundancy}
\noindent
The current Swarm deployment can support traditional erasure coding techniques on a per-chunk basis.
For example, by using Reed-Solomon coding~\cite{wicker1999reed}, a file consisting of $k$ chunks can be encoded using $n>k$ chunks so that any $k$-out-of-$n$ chunks can be used for retrieval.

Unfortunately, erasure coding the original file would not be sufficient, as this would only apply to the leaves of the Merkle tree, while the internal nodes would be vulnerable to data loss.

Our research revolves around finding better ways to add redundancy to the Merkle tree.
To that end, we have used several coding schemes and developed novel algorithms for use with Merkle trees.
We have evaluated our algorithms experimentally.
Next, we describe our setup and how we were able to evaluate our algorithms in a cluster of 1000 peers.

%% file: tex/evalutation.tex

\section{Experimental Setup}
\noindent
Our cluster consists of 30 machines running Ubuntu.
We use Helm~\cite{helm} and Kubernetes~\cite{kube} to distribute the load and quickly scale up to 1000 Swarm peers.
We avoid overloading the cluster machines to facilitate restarting the experiment if our algorithms cause a crash or get stuck while debugging.

Helm uses a single configuration file as input to a dozen configuration files during the cluster's initialization.
In the single configuration file, we specify parameters such as storage capacity, cluster placement, scaling, run-time parameters of Swarm, and much more.
We base our Helm scripts on those provided by the Ethereum Swarm organization~\cite{helm-charts}.

Peers running in Kubernetes are ephemeral by default, with limited options for persistent storage.
The before-mentioned Helm scripts only supported persistent storage options for cloud providers, but not for private clusters.
To provide persistent storage to the peers from the local SSD, we had to create a \emph{Persistent Volume} (PV).
Each peer has a dedicated PV, and each PV is linked to a physical location on the SSD.
We elegantly used the modulus operator in the Helm configuration file to create the link so that peer $y$'s data was allocated to a PV assigned to machine $y \mod 29$.

%% file: tex/dealingreplication.tex

\section{Dealing with Replication}
\label{sec:dealreplication}
\noindent
Each peer in Swarm is assigned a unique identifier in a Kademlia~\cite{maymounkov2002kademlia} overlay network, and
peers are grouped into neighborhoods based on the similarity of their identifiers.
Inside neighborhoods, peers attempt to replicate their data to each other.
Peers with the most similar identifier to a chunk is deemed responsible for persisting that chunk.

\begin{figure}[h]
	\setlength\abovecaptionskip{-0.5\baselineskip}
	\centering
	\includegraphics[width=\columnwidth]{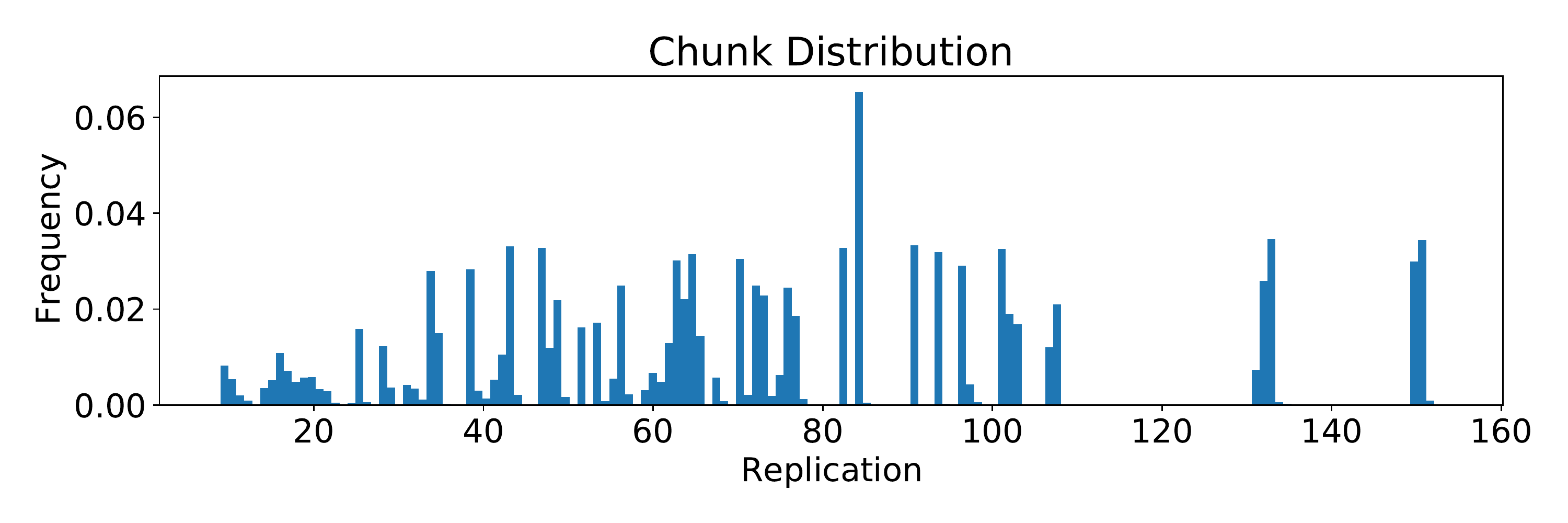}
	\caption{Chunk replication with 1000 Swarm peers.}
	\label{fig:nodesPerChunk}
\end{figure}

\begin{figure}[h]
	\setlength\abovecaptionskip{-0.5\baselineskip}
	\centering
	\includegraphics[width=\columnwidth]{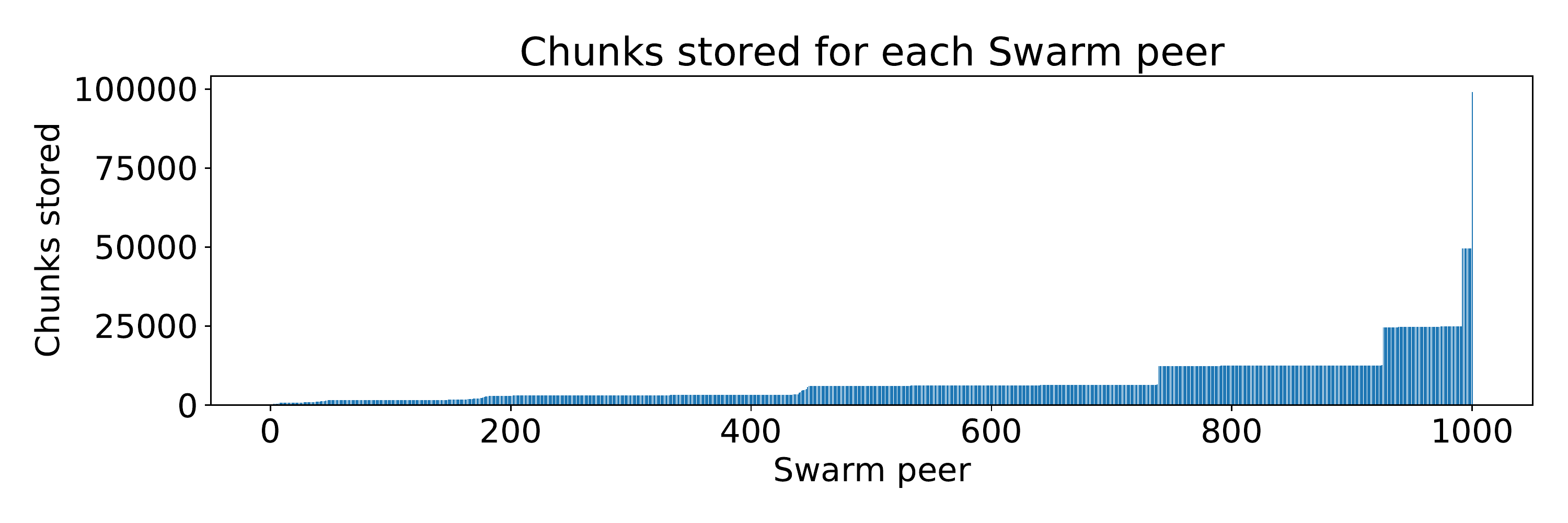}
	\caption{Chunks stored by each Swarm peer.}
	\label{fig:chunksPerNode}
\end{figure}

As peers have their own view of the overlay network, some ``superpeers'' may exist in multiple neighborhoods.
This results in some chunks being more replicated than others, and in our particular setup, this ranged from 9 replicas to 154 replicas.
\figref{fig:nodesPerChunk} shows how the chunks of a 100 MB file were replicated, and \figref{fig:chunksPerNode} shows how the chunks were distributed to the peers.
To ensure fair comparisons between algorithms, we need to replicate chunks exactly once.

\vspace{-.3cm}
\vspace{-.7cm}

\noindent
Swarm offers no tools to control each peer's storage, so the only way to achieve this was to develop our own tools.
Our first tool is named \emph{listchunks}~(1), and its main objective is to list the content address of all the chunks belonging to each file in our storage network.
The listchunks tool achieves this by merely appending each new chunk's content address it traverses from the root chunk when retrieving the file.

The second tool is \emph{bakedeletion}~(2), which is responsible for figuring out which chunks must be deleted from which peer.
It generates the deletion list by iteratively simulating the removal of chunks from peers while following four basic rules; (A) All peers must have some chunks, (B) Cardinality of unique chunks must be equal after deletion, (C) Peers can not get new chunks, (D) Replication factor for each chunk must be uniform.

To run experiments on different file sizes, we need to run tools (1) and (2) for each file we want to evaluate, as rule (A) must be respected for each file.
Thus, we need to aggregate all deletion lists, and this is done in the third tool \emph{combinestorage}~(3).

Finally, the aggregated deletion list is passed to \emph{deletechunks}~(4), which goes through the storage network, peer by peer, and deletes the mapped chunks.

%% file: tex/dealingsyncing.tex

\section{Dealing with Syncing}
\noindent
To ensure an identical system state across test iterations, we had to control the chunk distribution, or \emph{syncing}.
The \emph{push-syncing} and \emph{pull-syncing} can both be disabled by the command-line option \emph{no-sync} when starting the peer.

However, the syncing process that occurs when chunks are delivered through the Kademlia DHT can not be disabled.
Therefore we had to create our own procedures to create snapshots of the storage states and recover from a snapshot between each test iteration.

The snapshot of a peer is created by copying all the chunks stored by the peer immediately after we have made replication uniform for all chunks, as discussed in Section~\ref{sec:dealreplication}.
Before initiating the copying process, we must first terminate the peers to avoid data corruption.

To ensure an identical system state between test iterations, we (A) replace the local storage with the snapshots for each of the peers, (B) ensure that the \emph{no-sync} option is turned on and (C) ensure that each peer is well-connected.
When the test iteration concludes, we (D) shut down the peers, and for the next iteration, we continue from step (A) again.

We achieve step (A) by merely copying the snapshot to the peers using \emph{rsync}.
To reach sufficient connectivity required for step (C), the peers must first discover each other.
We monitor this process by periodically polling the inter-process communication file \emph{bzzd.ipc}.
As soon as the desired connectivity is reached, we can continue with the experimental evaluation, e.g., file availability is given peer failures.

%% file: tex/conclusion.tex
\section{Conclusion}
\noindent
In this paper, we have shared our experiences with running a 1000 peer cluster of Ethereum Swarm instances.
Without these tools to manage the peers, the configuration would be a nightmare scenario and ensuring they are all running correctly, even worse.
With the assistance of these tools, we can make fair comparisons of redundancy algorithms in a P2P storage system.